\documentclass[twocolumn,runningheads,natbib]{svjour2}
\smartqed  % flush right qed marks, e.g. at end of proof
\usepackage{graphicx}
\newcommand{\aap}{A\&A}

\newcommand{\apss}{Astrophysics and Space Science}
\newcommand{\apj}{ApJ}
\newcommand{\apjl}{ApJ}

\newcommand{\mnras}{MNRAS}
\newcommand{\pasp}{PASP}

\newcommand{\aj}{AJ}
\newcommand{\nat}{Nature}

\newcommand{\rxa}{RX\,J0420.0--5022}
\newcommand{\rxb}{RX\,J0720.4--3125}
\newcommand{\rxc}{RX\,J0806.4--4123}

\newcommand{\rxe}{RX\,J1605.3+3249}
\newcommand{\rxf}{RX\,J1856.4--3754}

\newcommand{\rbs}{RBS\,1223}
\newcommand{\rbt}{RBS\,1774}
\newcommand{\Halp}{H${\alpha}$}
\journalname{Astrophysics and Space Science}
\begin{document}

\title{The Magnificent Seven: Magnetic fields and surface temperature distributions
}


\author{Frank Haberl %\and
%        Second Author %etc.
}


\institute{F. Haberl \at
              Max-Planck-Institut f{\"u}r extraterrestrische Physik\\
	      Giessenbachstrasse, 85748 Garching, Germany\\
              Tel.: +49-89-300003320\\
              Fax:  +49-89-300003569\\
              \email{fwh@mpe.mpg.de}           %  \\
}

\date{Received: date / Accepted: date}

\maketitle

\begin{abstract}
Presently seven nearby radio-quiet isolated neutron stars discovered in 
ROSAT data and characterized by thermal X-ray spectra are known. They exhibit
very similar properties and despite intensive searches their number remained 
constant since 2001 which led to their name ``The Magnificent Seven''.
Five of the stars exhibit pulsations in their X-ray flux with periods in the
range of 3.4 s to 11.4 s. XMM-Newton observations revealed broad
absorption lines in the X-ray spectra which are interpreted as cyclotron
resonance absorption lines by protons or heavy ions and / or atomic transitions
shifted to X-ray energies by strong magnetic fields of the order of 
10$^{13}$ G. New XMM-Newton observations indicate more complex X-ray spectra
with multiple absorption lines.
Pulse-phase spectroscopy of the best studied pulsars \rxb\ and \rbs\ 
reveals variations in derived emission temperature and absorption line depth
with pulse phase. Moreover, \rxb\ shows long-term spectral changes which are
interpreted as due to free precession of the neutron star. Modeling of
the pulse profiles of \rxb\ and \rbs\ provides information about the 
surface temperature distribution of the neutron stars indicating hot 
polar caps which have different temperatures, different sizes and 
are probably not located in antipodal positions.

\end{abstract}

\section{Introduction}
\label{intro}
After the discovery of \rxf\ as an isolated neutron star in ROSAT data 
\citep{1996Natur.379..233W,1997Natur.389..358W}, six further objects with 
very similar properties were found. Their X-ray emission is characterized by
a soft, blackbody-like continuum which is little attenuated by photo-electric 
absorption by the interstellar medium, indicating small distances 
\cite[of the order of a few 100 pc,][]{2006NS...Posselt}.
For the brightest object, \rxf, this was soon confirmed by the parallax measurement 
\citep{2001ApJ...549..433W,2002ApJ...571..447K,2002ApJ...576L.145W}.
Including new HST observations the most recent revision of the parallax yields
a distance of 161$^{+18}_{-14}$ pc \citep{2006NS...vanKerkwijk}. 
A preliminary value for the parallax of \rxb\ is presented by \cite{2006NS...vanKerkwijk}
which corresponds to a distance of 330$^{+170}_{-80}$ pc.

No indication for a hard, non-thermal X-ray emission component was found in the 
X-ray spectra. Compared to normal radio pulsars also no strong radio emission is detected:
While for \rxb\ and \rxc\ deep radio observations failed to detect a radio 
counterpart \citep{2003MNRAS.340L..43J,2003ApJ...590.1008K}, for \rbs\ and \rbt\ the detection 
of weak radio emission was claimed \citep{2006ATel..798....1M,2006NS...Malofeev}. 
The constant X-ray flux on time scales of years, no obvious association with a 
supernova remnant and relatively high proper motion measurements for the 
three brightest objects suggest that we deal with a group of cooling neutron 
stars with ages of around 10$^6$ years. 
Recent reviews which summarize the observed properties of the seven thermally 
emitting INSs known today can be found in 
\cite{2000PASP..112..297T}, \cite{2001xase.conf..244M} and  
\cite{2004AdSpR..33..638H,2005fysx.conf...39H}.

\begin{table*}
\caption[]{X-ray and optical properties of the Magnificent Seven}
\centering
\label{tab-x-opt} 
\begin{tabular}{lcccccc}
\hline\noalign{\smallskip}
\multicolumn{1}{l}{Object} &
\multicolumn{1}{c}{kT} &
\multicolumn{1}{c}{Period} &
\multicolumn{1}{c}{Amplitude} &
\multicolumn{1}{c}{Optical} &
\multicolumn{1}{c}{PM} &
\multicolumn{1}{c}{Ref.} \\

\multicolumn{1}{l}{} &
\multicolumn{1}{c}{eV} &
\multicolumn{1}{c}{s} &
\multicolumn{1}{c}{\%} &
\multicolumn{1}{c}{mag} &
\multicolumn{1}{c}{mas/year} &
\multicolumn{1}{c}{} \\[3pt]
\tableheadseprule\noalign{\smallskip}
RX\,J0420.0--5022   &  44    &  3.45 & 13     & B = 26.6		&     & 1           \\
RX\,J0720.4--3125   &  85-95 &  8.39 & 8-15   & B = 26.6		&  97 & 2,3,4,5,6   \\
RX\,J0806.4--4123   &  96    & 11.37 &  6     & B $>$ 24		&     & 7,1         \\
RBS\,1223$^{(a)}$    &  86    & 10.31 & 18     & m$_{\rm 50ccd}$ = 28.6 &     & 8,9,10,11   \\
RX\,J1605.3+3249     &  96    & 6.88? & ?      & B = 27.2		& 145 & 12,13,14,15 \\
RX\,J1856.5--3754   &  62    & $-$   & $<$1.3 & B = 25.2		& 332 & 16,17,18,19 \\
RBS\,1774$^{(b)}$    & 102    & 9.44  &  4     & B $>$ 26		&     & 20,21,22    \\
\noalign{\smallskip}\hline\noalign{\smallskip}
\multicolumn{7}{l}{$^{(a)}$ = 1RXS\,J130848.6+212708} \\
\multicolumn{7}{l}{$^{(b)}$ = 1RXS\,J214303.7+065419} \\
\multicolumn{7}{l}{References:} \\
\multicolumn{7}{l}{
(1)  \cite{2004A&A...424..635H}
(2)  \cite{1997A&A...326..662H}
(3)  \cite{2001A&A...365L.302C}
} \\
\multicolumn{7}{l}{
(4)  \cite{2004A&A...419.1077H}
(5)  \cite{2004A&A...415L..31D}
(6)  \cite{2003A&A...408..323M}
} \\
\multicolumn{7}{l}{
(7)  \cite{2002A&A...391..571H}
(8)  \cite{1999A&A...341L..51S}
(9)  \cite{2002A&A...381...98H}
} \\
\multicolumn{7}{l}{
(10) \cite{2002ApJ...579L..29K}
(11) \cite{2003A&A...403L..19H}
(12) \cite{1999A&A...351..177M}
} \\
\multicolumn{7}{l}{
(13) \cite{2003ApJ...588L..33K}
(14) \cite{2004ApJ...608..432V}
(15) \cite{2005A&A...429..257M}
} \\
\multicolumn{7}{l}{
(16) \cite{1997Natur.389..358W}
(17) \cite{2002ApJ...576L.145W}
(18) \cite{2003A&A...399.1109B}
} \\
\multicolumn{7}{l}{
(19) \cite{2001A&A...378..986V}
(20) \cite{2001A&A...378L...5Z}
(21) \cite{2005ApJ...627..397Z}
(22) \cite{2006NS...Komarova}
} \\
\end{tabular} 
\end{table*}

The discovery of the seven neutron stars (which are often called the
Magnificent Seven, hereafter M7) with purely thermal X-ray spectra raised wide 
interest by theoreticians and observers as promising objects to learn about 
atmospheres and the internal structure of neutron stars \citep[e.g.][]{1997ApJ...476L..47P}.
The apparent absence of non-thermal processes which would hamper the analysis
of the X-ray emission allows a direct view onto the stellar surface.
Limits on the radius of the star \rxf\ were used to constrain the equation of state 
of neutron star matter \citep{2002ApJ...564..981P,2004NuPhS.132..560T}.
The ROSAT PSPC spectra with low energy resolution were consistent with 
Planckian energy distributions with blackbody 
temperatures kT in the range 40 -- 100 eV and little attenuation by 
interstellar absorption. More recent high resolution observations of
the two brightest objects \rxf\ \citep{2001A&A...379L..35B} and \rxb\ 
\citep{2001A&A...365L.298P,2002nsps.conf..273P,2003ApJ...590.1008K} 
were performed using the low
energy transmission grating (LETG) aboard Chandra and the reflection grating 
spectrometers (RGS) of XMM-New\-ton. In particular the high signal to noise 
LETG spectrum from a 500 ks Chandra observation of
\rxf\ did not reveal any significant deviation from a blackbody spectrum 
\citep{2001A&A...379L..35B,2003A&A...399.1109B}. 
The statistical quality and energy band coverage of the RGS and LETG spectra 
of \rxb\ are however insufficient to detect subtle narrow features
in the spectrum \citep{2001A&A...365L.298P}.

Four of the M7 exhibit clear pulsations in their X-ray flux, indicating the 
spin period of the neutron star (\rxa, \rxb, \rxc\ and \rbs) and there is
evidence that also \rbt\ is a pulsar \citep{2005ApJ...627..397Z}. A possible candidate period 
for \rxe\ needs future confirmation. A summary with blackbody temperatures
derived from (phase-averaged) EPIC-pn spectra, spin periods, amplitudes 
for the flux modulation,
optical brightness and measured proper motions is given in Table~\ref{tab-x-opt}
together with key references for each object.
Despite extensive searches in ROSAT data 
\citep[e.g. ][]{2003ApJ...598..458R,2005A&A...444...69C,2006AJ....131.1740A} 
no further neutron star with similar characteristic properties was found.

The first significant deviations from a blackbody spectrum were reported 
by \cite{2003A&A...403L..19H} from EPIC spectra of \rbs. A broad absorption 
feature near 300 eV was interpreted as proton cyclotron resonance absorption.
Apart from this absorption line with an equivalent width of 160 eV
the X-ray spectrum of \rbs\ is blackbody-like. \cite{2003A&A...403L..19H} 
also suggested that changes in the soft part of the X-ray spectrum of the 
pulsar \rxb\ with pulse phase reported by \cite{2001A&A...365L.302C} are 
caused by variable cyclotron absorption \citep{2004A&A...419.1077H}. 
Cyclotron resonance absorption features in the 0.1--1.0 keV band are expected 
in spectra from strongly magnetized neutron stars with field strengths in the range 
of $10^{10} - 10^{11}$ G or $2 \times 10^{13} - 2 \times 10^{14}$ G if 
caused by electrons or protons, respectively 
\citep[see e.g.][]{2001ApJ...560..384Z,2002nsps.conf..263Z}. 
Variation of the magnetic field strength over the neutron star surface 
leads to a broadening of the line \citep{2004ApJ...607..420H}.

The relatively long spin periods in the range of 3.45~s $-$ 11.37~s 
(the majority of radio pulsars has periods less than 1~s) were the first 
indicator for strong magnetic fields. If the stars were born
with millisecond spin periods, age estimates from neutron star cooling curves
\citep{2006NuPhA...Page} and proper motions 
\cite[tracing the neutron star back to a likely birth place,][]{2006NS...Motch}
of typically 10$^6$ years require B fields of the order of 10$^{13}$ G to 
decelerate the rotation of the stars to their current periods.
This was recently confirmed by the accurate determination of the pulse period
derivative $\dot P$ of $0.698 \times 10^{-13}$ s s$^{-1}$ and 
$1.120 \times 10^{-13}$ s s$^{-1}$ for \rxb\ and \rbs, respectively 
\citep{2005ApJ...628L..45K,2005ApJ...635L..65K}. In the magnetic dipole braking
model this yields characteristic ages of $1.9 \times 10^{6}$ years and 
$1.5 \times 10^{6}$ years and B field strengths of  $2.4 \times 10^{13}$ G and 
$3.4 \times 10^{13}$ G, respectively. Such strong fields supports the picture 
in which the broad absorption lines -- if due to cyclotron resonance -- at least for 
\rxb\ and \rbs\ are caused by protons.

\begin{figure*}
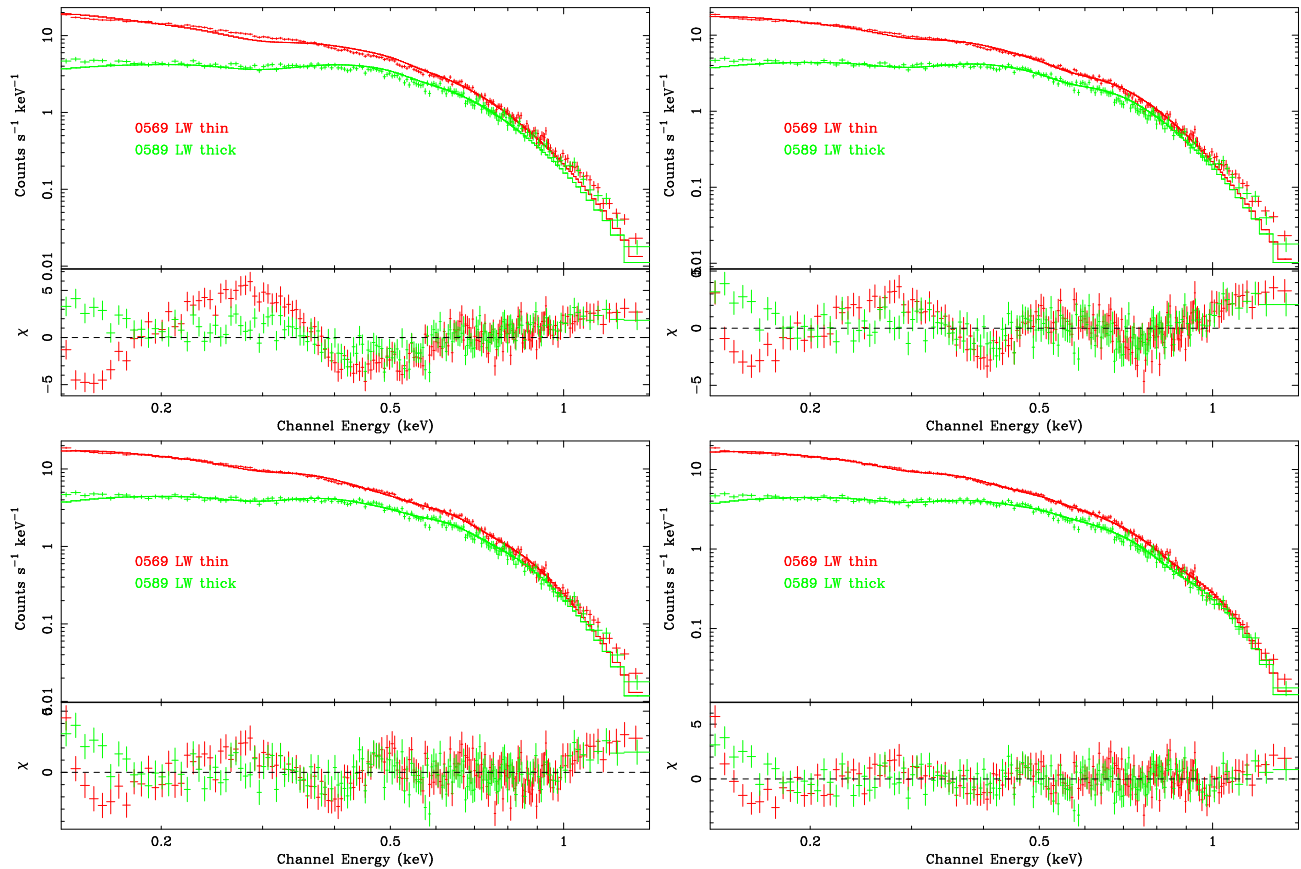

\centering
  \hbox{
  \resizebox{0.485\hsize}{!}{\includegraphics[clip=,angle=-90]{p015736_pn0_bb.ps}}
  \resizebox{0.485\hsize}{!}{\includegraphics[clip=,angle=-90]{p015736_pn0_bbgauss.ps}}
  }
  \hbox{
  \resizebox{0.485\hsize}{!}{\includegraphics[clip=,angle=-90]{p015736_pn0_bb2gauss.ps}}
  \resizebox{0.485\hsize}{!}{\includegraphics[clip=,angle=-90]{p015736_pn0_bb3gauss.ps}}
  }
\caption{Modeling of the EPIC-pn spectra of \rxe. From top left to bottom right
the number of absorption lines increases from zero to three. As continuum a
blackbody model with interstellar absorption is used.}
\label{fig-spectra}
\end{figure*}

\section{Pulse-phase averaged X-ray spectra}

Significant deviations from the absorbed blackbody spectrum were reported
from XMM-Newton EPIC observations of \rbs\ \citep{2003A&A...403L..19H} and 
\rxb\ \citep{2004A&A...419.1077H} and from XMM-Newton RGS  
spectra of \rxe\ \citep{2004ApJ...608..432V}. Also non-magnetic neutron star 
atmosphere models \cite[e.g.][]{2002A&A...386.1001G,2002nsps.conf..263Z}
fail to reproduce the spectra. Iron and solar mixture atmospheres
cause too many absorption features in particular at energies between 0.5 and 1.0 keV 
which are not seen in the measured spectra. On the other 
hand the spectrum of a pure hydrogen model is similar in shape to that of a blackbody
but results in a much lower effective temperature which would predict a far too 
high optical flux \citep[see][]{1996ApJ...472L..33P}. The modeling of the X-ray 
spectra was significantly improved by including a broad absorption line 
with Gaussian shape. Line energies around 300 eV were found for \rxb\ and \rbs\
from the medium energy resolution EPIC spectra, while a line with higher 
energy of 450 -- 480 eV was discovered in the high energy resolution RGS 
spectrum of \rxe. 

\begin{table*}
\caption[]{Model fits to the EPIC-pn spectra of \rxe}
\centering
\label{tab-spectra} 
\begin{tabular}{cccccccccccc}
\hline\noalign{\smallskip}
\multicolumn{1}{c}{$\mathrm{N_H}$} &
\multicolumn{1}{c}{kT} &
\multicolumn{1}{c}{E$_{1}$} &
\multicolumn{1}{c}{E$_{2}$} &
\multicolumn{1}{c}{E$_{3}$} &
\multicolumn{1}{c}{F$_{1}$} &
\multicolumn{1}{c}{F$_{2}$} &
\multicolumn{1}{c}{F$_{3}$} &
\multicolumn{1}{c}{EW$_{1}$} &
\multicolumn{1}{c}{EW$_{2}$} &
\multicolumn{1}{c}{EW$_{3}$} &
\multicolumn{1}{c}{$\chi_{red}^2$} \\

\multicolumn{1}{c}{10$^{20}$ cm$^{-2}$} &
\multicolumn{1}{c}{eV} &
\multicolumn{1}{c}{eV} &
\multicolumn{1}{c}{eV} &
\multicolumn{1}{c}{eV} &
\multicolumn{3}{c}{$10^{-4}$ photons cm$^{-2}$ s$^{-1}$} &
\multicolumn{1}{c}{eV} &
\multicolumn{1}{c}{eV} &
\multicolumn{1}{c}{eV} &
\multicolumn{1}{c}{} \\[3pt]
\tableheadseprule\noalign{\smallskip}
0.15          & 97           & --        & --         & --         & --        & --           & -- & & & & 4.38 \\
1.1           & 92           & 445       & --         & --         & $-67$     & --           & -- & & & & 2.39 \\
1.4           & 92           & 440       & 1.5E$_{1}$ & --         & $-75$     & $-43$        & -- & & & & 1.75 \\
2.04$\pm$0.04 & 91.6$\pm$0.3 & 403$\pm$2 & 589$\pm$4  & 780$\pm$24 & $-43\pm$1 & $-8.0\pm$0.8 & $-1.6\pm$0.4 & 96 & 76 & 67 & 1.39 \\
\noalign{\smallskip}\hline\noalign{\smallskip}
\end{tabular}
\end{table*}

\begin{figure}
\centering
  \resizebox{\hsize}{!}{\includegraphics[clip=,angle=-90]{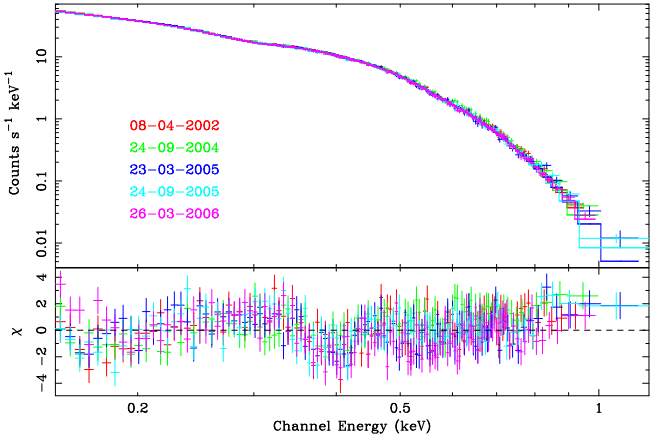}}
\caption{EPIC-pn spectra of \rxf\ obtained from five observations between 2002 and 2006 
         with the same instrumental setup (small window readout mode and thin filter). 
	 A joint fit with an absorbed blackbody model
	 yields a reduced $\chi^2$ of 1.55 and demonstrates the stability of detector 
	 and source to better than 1\%. The dip in the residuals around 0.4 keV is a remaining
	 calibration problem.}
\label{fig-rxf}
\end{figure}

\begin{figure}
\centering
  \resizebox{\hsize}{!}{\includegraphics[clip=,angle=-90]{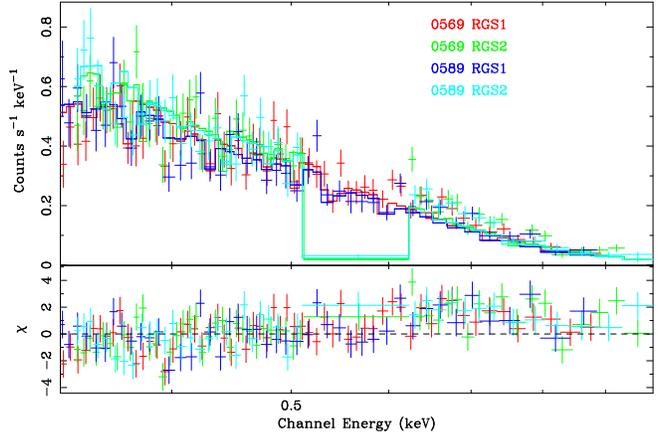}}
\caption{RGS spectra of \rxe\ from the two observations in 2003, January 17 
and February 26 (satellite revolutions 569 and 589). The histograms represent the 
best fit EPIC-pn model with three absorption lines allowing only a re-normalization
to account for possible cross-calibration problems.}
\label{fig-rgsspectra}
\end{figure}

Spectral fits to the EPIC-pn spectra of the third brightest M7 star 
\rxe\ obtained in 2003, January 17 and February 26 (satellite revolutions 
569 and 589) also considerably improve when
adding a Gaussian absorption line to the absorbed blackbody model (see 
Table~\ref{tab-spectra} and Fig.~\ref{fig-spectra}). However, the 
quality of the fit with a reduced $\chi^2$ of 2.39 (all XMM-Newton spectra 
presented here were obtained using the analysis software SAS version 6.5) 
is still not acceptable and worse than typical values around 1.5 one finds 
for joint fits to the available EPIC-pn spectra of 
\rxf\ (see Fig.~\ref{fig-rxf}) or 
\rxb. Adding more lines further improves the fit quality and with three absorption
lines an acceptable $\chi^2$ is found. The best fit parameters for the models with 
different number of absorption lines are listed in Table~\ref{tab-spectra}.
The width of the lines was fit as a single common parameter in the case
of multiple lines ($\sigma$ = 87 eV for the model with three lines).
A remarkable result is the ratio of the line energies for the model with three lines.
With E$_2$/E$_1$ = 1.46$\pm$0.02 and E$_3$/E$_1$ = 1.94$\pm$0.06 the line energy ratios
are consistent with 1:1.5:2. It should also be noted that the depth of the lines 
decreases with line energy by a factor of about 5 (in terms of absorbed line
photon flux) from one line to the next.
A fit with the model with three absorption lines to the RGS spectra (only allowing
a re-normalization factor between the instruments) shows that also the RGS spectra 
are consistent with that model (Fig.~\ref{fig-rgsspectra}).

Similar to \rxe, new XMM-Newton observations also indicate a more complex X-ray spectrum for \rbs.
With a total exposure time of more than 100~ks the increased statistical quality of the
EPIC-pn spectra requires two absorption lines for an acceptable fit \citep{2006NS...Schwope}. 
With 230 eV and 460 eV the line energies are a factor of two apart.
Also, in the case of \rxc\ a simple absorbed
blackbody model yields an unacceptable fit to the EPIC-pn spectra
\citep{2004A&A...424..635H}. A model with one line
results in $\chi^2_{red} = 1.39$ which is formally acceptable. Adding another line with
an energy twice that of the first one, further improves the fit to $\chi^2_{red} = 1.05$, but
more higher quality data is required to confirm the significance of this result.

\begin{table*}
\caption[]{Magnetic field estimates}
\centering
\label{tab-Bfields} 
\begin{tabular}{lcccc}
\hline\noalign{\smallskip}
\multicolumn{1}{l}{Object} &
\multicolumn{1}{c}{dP/dt} &
\multicolumn{1}{c}{E$_{\rm cyc}$} &
\multicolumn{1}{c}{B$_{\rm db}$} &
\multicolumn{1}{c}{B$_{\rm cyc}$} \\

\multicolumn{1}{l}{} &
\multicolumn{1}{c}{10$^{-13}$ ss$^{-1}$} &
\multicolumn{1}{c}{eV} &
\multicolumn{1}{c}{10$^{13}$ G} &
\multicolumn{1}{c}{10$^{13}$ G} \\[3pt]
\tableheadseprule\noalign{\smallskip}
RX\,J0420.0$-$5022   &  $<$92         & 330             & $<$18           & 6.6       \\
RX\,J0720.4$-$3125   &  0.698(2)      & 280             & 2.4             & 5.6       \\
RX\,J0806.4$-$4123   &  $<$18         & 430/306$^{(a)}$ & $<$14           & 8.6/6.1   \\
RBS\,1223            &  1.120(3)      & 300/230$^{(a)}$ & 3.4             & 6.0/4.6   \\
RX\,J1605.3+3249     &                & 450/400$^{(b)}$ &                 & 9/8       \\
RX\,J1856.5$-$3754   &                & $-$             & $\sim$1$^{(c)}$ & 	    \\
RBS\,1774            &  $<$60$^{(d)}$ & 750             & $<$24$^{(d)}$   & 15        \\
\noalign{\smallskip}\hline\noalign{\smallskip}
\multicolumn{5}{l}{$^{(a)}$ Spectral fit with single / two lines} \\
\multicolumn{5}{l}{$^{(b)}$ With single line / three lines at 400 eV, 600 eV and 800 eV} \\
\multicolumn{5}{l}{$^{(c)}$ Estimate from \Halp\ nebula assuming that it is powered by magnetic} \\
\multicolumn{5}{l}{\hspace{4mm} dipole breaking \citep{2002ApJ...571..447K,2002ApJ...580.1043B,2004NuPhS.132..560T}} \\
\multicolumn{5}{l}{$^{(d)}$ Radio detection: \cite{2006ATel..798....1M}} \\
\end{tabular}
\end{table*}

Table~\ref{tab-Bfields} summarizes the magnetic field estimates for the M7 neutron stars 
utilizing either the magnetic dipole braking model for the pulsars with measured pulse 
period derivative $\dot P$ (B = $3.2 \times 10^{19} (P \times \dot P)^{1/2}$)
or assuming that the absorption line is due to cyclotron resonance by protons 
(B = $1.6 \times 10^{11} E(eV)/(1-2GM/c^2R)^{1/2}$). If multiple absorption lines 
were included in the spectral modeling, the lowest line energy was used for the estimate 
of the magnetic field.

The brightest M7 star, \rxf, does not show pulsations with a very low amplitude limit 
of 1.3\% \citep{2002ApJ...570L..75R,2003A&A...399.1109B}. However, 
\cite{2001A&A...380..221V} discovered
the existence of a \Halp\ nebula around the star with cometary-like morphology aligned
with the direction of its proper motion. Assuming that the nebula is powered by magnetic 
dipole braking and an age of the neutron star of $\sim$5 $\times 10^{5}$ years which is 
inferred from its proper motion and the distance to the likely birth place, a magnetic 
field strength of $\sim$1 $\times$ 10$^{13}$ G is derived 
\citep{2002ApJ...571..447K,2002ApJ...580.1043B,2004NuPhS.132..560T}.
Such a field strength is consistent with the non-detection of a proton cyclotron
feature which would have an energy below the sensitive range of X-ray instruments.

In two cases (\rxb\ and \rbs) accurate $\dot P$ measurements exist. For both stars also 
absorption lines were found in their X-ray spectra which allows a comparison of 
B$_{\rm db}$ (field estimate from dipole braking) and B$_{\rm cyc}$ (field estimate from
proton cyclotron resonance). The ratio B$_{\rm cyc}$/B$_{\rm db}$ is 2.33 for \rxb\ and 
1.35 for \rbs\ (for the spectral model with two lines). A similar low ratio for \rxb\ would
be obtained if a second (lower-energy) line around 140 eV exists. Unfortunately, 
the detection of a line at such low energies is outside the capability of the XMM-Newton 
instruments.

\section{Neutron star surface temperature distributions}

The first XMM-Newton observations of \rxb\ revealed spectral changes with pulse phase
expressed as hardness ratio variations \citep{2001A&A...365L.302C}. Using polar cap models 
first constraints on the polar cap sizes and viewing geometries could be derived.
The pulse profile in the 0.12 keV to 1.2 keV energy band was found to be approximately 
sinusoidal with a peak-to-peak amplitude of about 15\% with the hardness ratio being 
softest slightly before flux maximum. Spectra from different pulse phases indicated 
a temperature variation and a change in the low-energy attenuation, which - when modeled 
with a simple absorbed blackbody spectrum - appears as variable absorption column density. 
Including more data \cite{2004A&A...419.1077H} interpreted this as broad absorption line 
with variable depth. Pulse phase spectroscopy of \rbs\ \citep{2005A&A...441..597S} also 
revealed that changes in temperature and absorption line depth are responsible for the 
spectral variations with pulse phase. The double-humped X-ray light curves of \rbs\ in 
different energy bands were modeled by \cite{2005A&A...441..597S} assuming Planckian 
radiation from a neutron star surface with inhomogeneous temperature distribution. Again,
a simple model with two hot spots with temperatures T$_1^\infty$ = 92 eV and 
T$_2^\infty$ = 84 eV and full angular sizes of 8$^\circ$ and 10$^\circ$ which are separated 
by $\sim160^\circ$ can reproduce the data. A more physical temperature
distribution based on the crustal field models by \cite{2004A&A...426..267G} %Geppert et al. 2004
which can produce relatively strong temperature gradients from the magnetic poles to 
the equator was equally successful.

\begin{figure*}
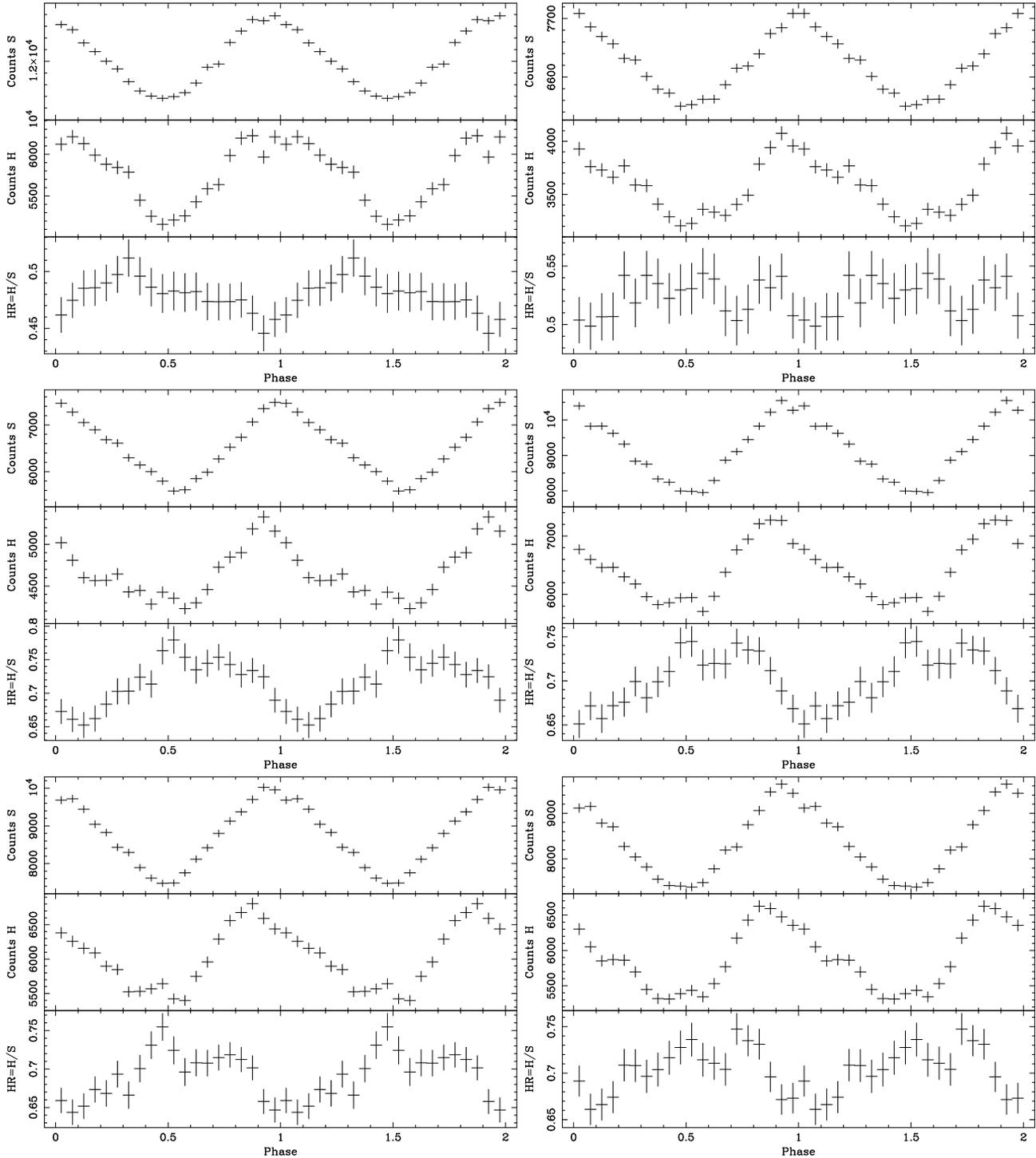

\centering
  \hbox{
  \resizebox{0.485\hsize}{!}{\includegraphics[clip=,angle=-90]{P0124100101PNS003PIEVLI0000_sc_hr_phase.ps}}
  \resizebox{0.485\hsize}{!}{\includegraphics[clip=,angle=-90]{P0156960201PNS003PIEVLI0000_sc_hr_phase.ps}}
  }
  \hbox{
  \resizebox{0.485\hsize}{!}{\includegraphics[clip=,angle=-90]{P0164560501PNS001PIEVLI0000_sc_hr_phase.ps}}
  \resizebox{0.485\hsize}{!}{\includegraphics[clip=,angle=-90]{P0300520201PNS003PIEVLI0000_sc_hr_phase.ps}}
  }
  \hbox{
  \resizebox{0.485\hsize}{!}{\includegraphics[clip=,angle=-90]{P0300520301PNS003PIEVLI0000_sc_hr_phase.ps}}
  \resizebox{0.485\hsize}{!}{\includegraphics[clip=,angle=-90]{P0311590101PNS003PIEVLI0000_sc_hr_phase.ps}}
  }
\caption{Folded EPIC-pn light curves of \rxb\ in two different energy bands (soft S: 120 eV - 400 eV;
         hard H: 400 eV - 1 keV) together with the hardness ratio. Different observations, all performed
	 with the same instrumental setup (full-frame mode with thin filter), are shown from 
	 top left to bottom right: 2000 May 13, 2002 Nov. 6, 2004 May 22, 2005 April 28, 
	 2005 Sep. 23 and 2005 Nov. 12.
	 For a detailed summary of all XMM-Newton observation in the years 2000 to 2005 see Table~1 in 
	 \cite{2006A&A...451L..17H}. Pulse phases were calculated using the X-ray timing ephemerides 
	 (``All Data'' solution) from \cite{2005ApJ...628L..45K}.}
\label{fig-lcurves}
\end{figure*}

The neutron star \rxb\ is unique among the M7 as it shows a gradual change of its 
X-ray spectrum over years which is accompanied by an energy-dependent change in the 
pulse profile \citep{2004A&A...415L..31D,2004ApJ...609L..75V}. 
Figure~\ref{fig-lcurves} shows folded light curves together with hardness ratios for six 
different epochs. They were derived from observations with identical instrumental setup
to avoid systematic differences between different filters and CCD readout modes.
As was found by \cite{2004A&A...415L..31D} the pulse profile became deeper with time
with a pulsed fraction of $\sim$8\% in the year 2000, increasing to $\sim$15\% by the 
end of 2003.
Moreover, the spectra became considerable harder during these years as can be seen in the 
increase of the average hardness ratio in Fig.~\ref{fig-lcurves}.
\cite{2006A&A...451L..17H} found first evidence for a trend reversal of the 
spectral evolution and interpret the involved period of 7.1$\pm$0.5 years as 
precession period of the neutron star. 

\begin{figure*}
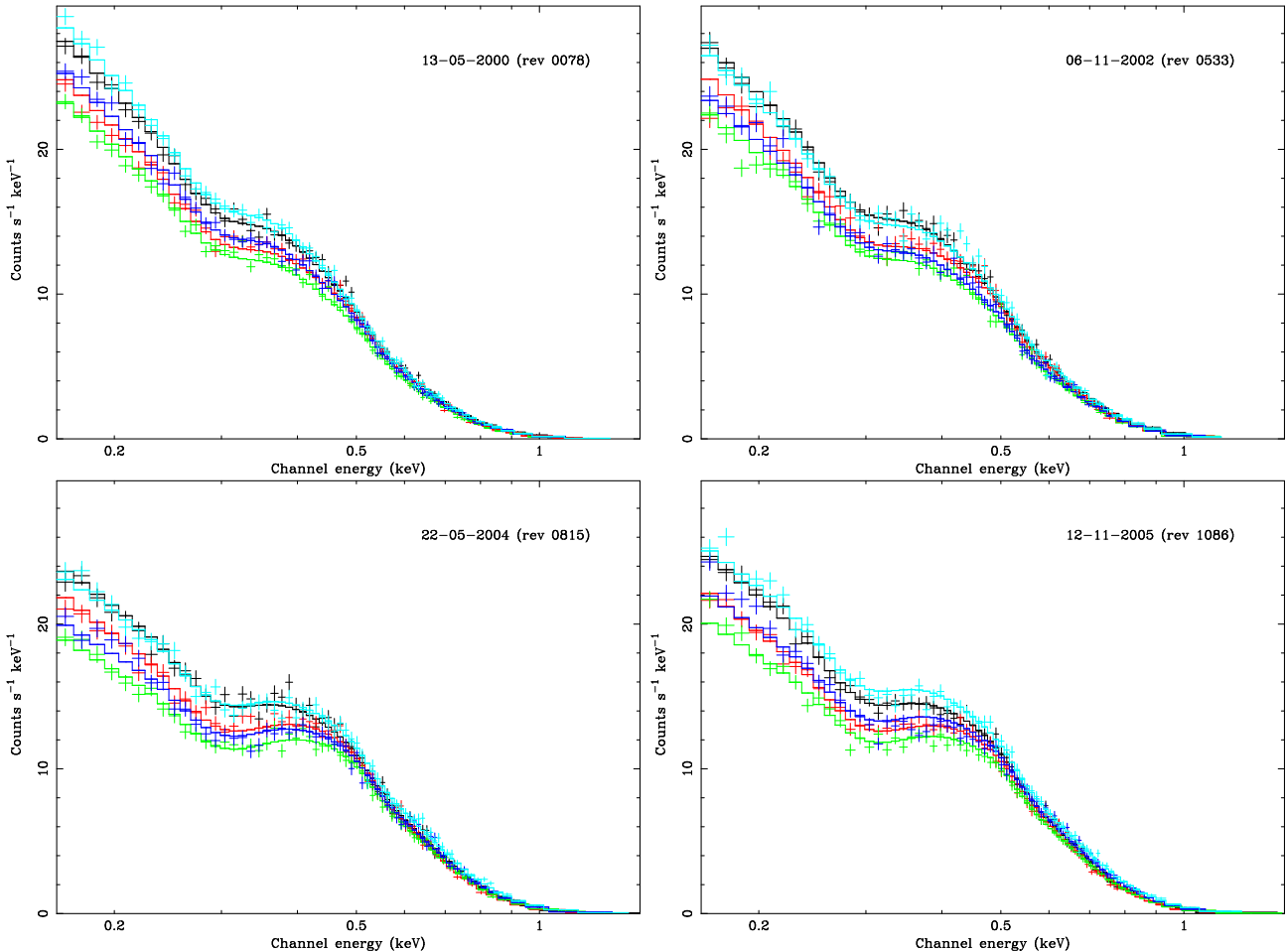

\centering
  \hbox{
  \resizebox{0.485\hsize}{!}{\includegraphics[clip=,angle=-90]{allffthin_phase5_o1.ps}}
  \resizebox{0.485\hsize}{!}{\includegraphics[clip=,angle=-90]{allffthin_phase5_o2.ps}}
  }
  \hbox{
  \resizebox{0.485\hsize}{!}{\includegraphics[clip=,angle=-90]{allffthin_phase5_o4.ps}}
  \resizebox{0.485\hsize}{!}{\includegraphics[clip=,angle=-90]{allffthin_phase5_o7.ps}}
  }
\caption{Pulse-phase resolved EPIC-pn spectra of \rxb\ from four different observations.
         During the pulse the spectra vary as is shown by the different colours used for
	 different phases (0.0-0.2: black, 0.2-0.4: red, 0.4-0.6: green, 0.6-0.8: blue, 
	 0.8-1.0: light blue; with phase 0.0 defined as intensity maximum as in 
	 Fig.~\ref{fig-lcurves}). The intensity scale is linear to better resolve the
	 changes at low energies.}
\label{fig-phasespectra}
\end{figure*}

\begin{figure}
\centering
  \resizebox{\hsize}{!}{\includegraphics[clip=,angle=-90]{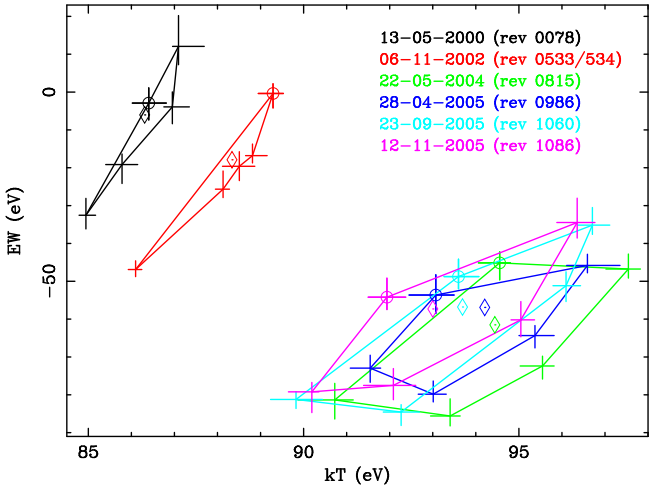}}
\caption{Equivalent width of the absorption line vs. temperature kT derived from
   the EPIC-pn full-frame mode observations with thin filter. Diamonds denote the values derived
   from the phase-averaged spectra. During the pulse the parameters evolve
   counter-clockwise, the circle marks phase 0.0-0.2. This figure is a colour representation
   of Fig.~4 in \cite{2006A&A...451L..17H}.}
\label{fig-eqwkt}
\end{figure}

The overall hardening of the spectra until 2004 and the trend-reversal can also be seen
from the pulse-phase resolved spectra which are shown in Fig.~\ref{fig-phasespectra}
for four different XMM-Newton observations. Both, pulse-phase variations and long-term 
spectral changes can be modeled by variations in the blackbody temperature and absorption 
line depth \citep{2006A&A...451L..17H}. 
In Fig.~\ref{fig-eqwkt} the line equivalent width is plotted versus temperature kT. 
The temperature variation ($\sim$2.5 eV) was smaller during the first observations 
and increased to $\sim$6 eV, almost as large as the long-term change of $\sim$8 eV seen
in the phase-averaged spectra. This supports the idea that the temperature changes are 
caused by geometrical effects and different areas on the neutron star surface come
into our view. In contrast, the amplitude in the equivalent width variation is $\sim$40 eV
and did not change much over the years. The relatively large variation in the line depth
with pulse phase and the strong increase over the years indicates that the hotter polar cap 
which came into (better) view -- possibly due to precession -- plays the major role in the line
absorption. More detailed analyses are required to see if this is mainly a temperature effect
(for higher temperatures atomic line transitions are expected to be reduced because 
of a higher ionization degree) or a viewing effect due to the dependence on the angle 
between the direction of radiation propagation and magnetic field lines.

\begin{figure}
\centering
  \resizebox{\hsize}{!}{\includegraphics[clip=,angle=-90]{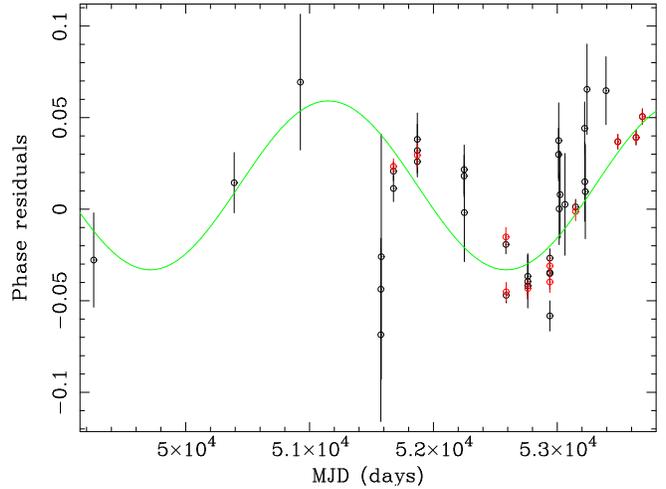}}
\caption{Phase residuals for \rxb. The black data points are reproduced from 
         \cite{2005ApJ...628L..45K} while red points are from the analysis of 
         \cite{2006A&A...451L..17H}, which includes a re-analysis of published EPIC-pn results 
         for a consistency check as well as three new data sets (the last three points).}
\label{fig-phaseres}
\end{figure}

Since the spectral observations do not yet cover a complete precession period 
\cite{2006A&A...451L..17H} also performed a preliminary analysis
of published and new pulse timing residuals (from three new XMM-Newton observations).
This includes archival ROSAT, Chandra and XMM-Newton observations which are distributed 
over a total of $\sim$12 years 
\citep[see ][]{2002ApJ...570L..79K,2002MNRAS.334..345Z,2004MNRAS.351.1099C,2005ApJ...628L..45K}. 
A sinusoidal fit to the phase residuals yields a period of 7.7$\pm$0.6 years
(Fig.~\ref{fig-phaseres}) consistent with that derived from the spectral analysis. 
In a similar analysis \cite{2006NS...vanKerkwijk} 
infer a shorter period of 4.3 years. Apart from using new Chandra data instead of the
new XMM-observations they also used different phase residuals for the early ROSAT 
observations (which are subject to cycle count ambiguity) compared to the original values
published in \cite{2005ApJ...628L..45K}. The reason for the different periods might
be due to the energy dependence 
of the pulse profile which can lead to systematic differences between instruments
with different spectral response and/or the deviations from a sinusoidal shape 
of the light curves (which moreover evolve with time) as it was assumed in the 
analysis of \cite{2005ApJ...628L..45K}. Further monitoring of \rxb\ and a more 
sophisticated analysis and modeling of the data are required to confirm the 
precession model and further constrain its parameters. This can provide independent 
information about the interior of the neutron star and its distortion from a 
spherical shape
$\epsilon=(\mathrm{I}_3-\mathrm{I}_1)/\mathrm{I}_1=\mathrm{P_{spin}}/\mathrm{P_{prec}}\approx 4\times 10^{-8}$
which is larger than that reported for radio pulsars 
\citep[e.g.][]{2001MNRAS.324..811J,2006MNRAS.365..653A} 
but smaller than that for Her\,X-1 \citep{1986ApJ...300L..63T,2000Ketsaris}. 

\cite{2006A&A...451L..17H} used a polar cap model similar to that applied by 
\cite{2005A&A...441..597S} to \rbs. Including free precession of the neutron star
to explain the long-term spectral variations of \rxb\ they can at least qualitatively 
reproduce the variations in the X-ray spectra, changes in the pulsed fraction, 
shape of the light curve and the phase-lag between soft and hard energy bands.
Like for \rbs, where the intensity peaks in the pulse profile are not separated by 
0.5 in spin phase, the spots on the surface of \rxb\ are probably also not located 
exactly in antipodal positions. This can explain the observed evolution of the light 
curve  and hardness ratio with precession phase.

\section{Discussion}

\begin{figure}
\centering
  \resizebox{\hsize}{!}{\includegraphics[clip=,angle=90]{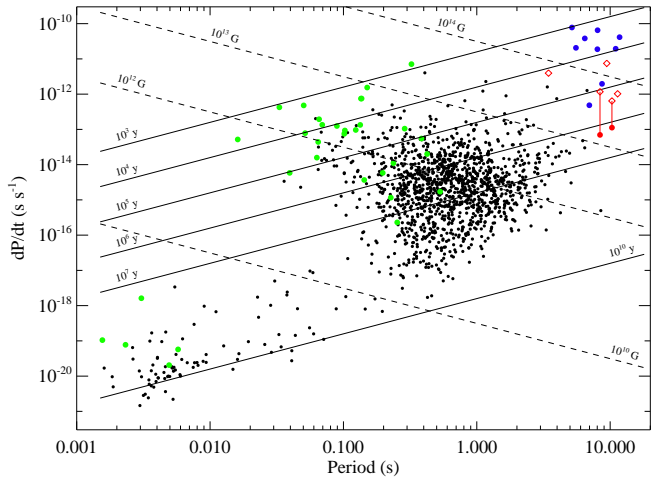}}
\caption{$P - \dot P$ diagram for radio pulsars (black dots), magnetars (blue dots)
   and the Magnificent Seven. Red dots show the direct $P$ and $\dot P$ measurements 
   for \rxb\ and \rbs, while the open diamonds mark the $\dot P$ values expected 
   from the magnetic field estimates assuming proton resonance as origin of the
   absorption lines. For \rxb\ and \rbs\ both field estimates are connected with 
   a vertical line. Marked in green are radio pulsars which are also detected at high
   energies (incomplete for the millisecond pulsars).}
\label{fig-ppdot}
\end{figure}

The new results from optical and X-ray observations of the Magnificent Seven,
thermally emitting isolated neutron stars, strongly support the model of cooling 
neutron stars with ages around a million years. Large proper motions measured 
for the three brightest objects make accretion from the interstellar medium to 
re-heat the neutron star very inefficient. Absorption features in the X-ray 
spectra of most of the Magnificent Seven and first measurements of the spin period
derivative from the two pulsars \rxb\ and \rbs\ consistently point to magnetic field
strengths of the order of 10$^{13}$ G to 10$^{14}$ G. This places them
at the long spin period and high magnetic field end of the radio pulsar distribution.
However, the $\dot P$ measurements still indicate field strengths below those of the 
magnetars (Fig.~\ref{fig-ppdot}). The discovery of a few radio pulsars with similar 
magnetic field strengths and long periods 
\citep{2000ApJ...541..367C,2002MNRAS.335..275M,2003ApJ...591L.135M} shows that radio 
emission can still occur at inferred field strengths close to the ``quantum critical field"
$B_{cr}= m_e^2c^3/e\hbar \simeq 4.4\times 10^{13}$ G. Therefore it remains unclear if 
the M7 exhibit no radio emission at all or if we do not detect them because their radio 
beam is very narrow due to their large light cylinder radius and therefore 
does not cross the Earth.

In a $\sim$10$^{13}$ G magnetic field cyclotron resonance lines at soft X-ray energies
are expected to be caused by protons or highly ionized atoms of heavy elements. 
The ratios of line strengths in consecutive harmonic lines scales with E$_{\rm cyc}/(mc^2)$ 
\citep{1980Ap&SS..73...33P} and for the involved high particle masses no harmonic lines
should be detectable. There is now more and more evidence for additional absorption 
lines in the X-ray spectra of the Magnificent Seven. An explanation for the
additional (or even all) lines can be atomic bound-bound or bound-free transitions. In high B 
fields atomic orbitals are distorted into a cylindrical shape and the electron 
energy levels are similar to Landau states, with binding energies of atoms 
strongly increased. For hydrogen in a magnetic field of the order of 10$^{13}$ G 
the strongest atomic transition is expected at energy 
E/eV $\approx$ 75(1+0.13ln(B$_{13}$))+63B$_{13}$,
with B$_{13}$ = B/10$^{13}$ G \citep{2002nsps.conf..263Z}. 
For the line energies found in the spectra of thermal isolated neutron stars this
requires similar field strengths to those derived assuming cyclotron absorption.
Atomic line transitions are expected to be less prominent at higher
temperatures because of a higher ionization degree \citep{2002nsps.conf..263Z}. 
For a more detailed discussion of expected line features that can be produced
in a hydrogen atmosphere see \cite{2006NS...vanKerkwijk}. It is also not clear how 
important the contribution of heavier elements in the atmosphere is because of the 
strong gravitational stratification forces \citep{2003astroph...Mori}.
Further, the remarkable harmonic-like energy spacing of multiple lines needs to 
be understood. Studying their pulse phase dependence which can probably best be done 
for \rbs\ might shed light on this.

The modeling of pulse profiles in different X-ray energy bands is a powerful tool 
to unveil the surface temperature distribution of the neutron stars \citep{2006MNRAS.366..727Z}.
First studies of \rxb\ and \rbs\ yielded constraints on the polar cap geometry.
It is remarkable that for both stars asymmetric configurations of the hot polar caps
were inferred. In both cases different temperatures and sizes of the hot spots
are found with the smaller spot being the hotter one. Probably, also in both stars 
the spots are not located in exactly antipodal positions. This could point to
magnetic field configurations of an off-centered dipole or the involvement of
higher-order multi-pole components. Studies of the effects of a strong magnetic
field on the temperature distribution in the neutron star crust show that
configurations which include dipolar poloidal and toroidal components can indeed 
reproduce the observed temperature distributions 
\citep{2004A&A...426..267G,2005astroph...Geppert}.

\begin{acknowledgements}
The XMM-Newton project is an ESA Science Mission with instruments
and contributions directly funded by ESA Member States and the
USA (NASA). The XMM-Newton project is supported by the
Bundesministerium f\"ur Wirtschaft und Technologie/Deutsches Zentrum
f\"ur Luft- und Raumfahrt (BMWI/DLR, FKZ 50 OX 0001), the Max-Planck
Society and the Heidenhain-Stif\-tung.
\end{acknowledgements}

% BibTeX users please use
%\bibliographystyle{spmpsci}
%%\bibliographystyle{aa}
%%\bibliography{ins,general,myrefereed,myunrefereed}   % name your BibTeX data base

%\begin{thebibliography}{3}
%\bibitem{Ref1}
%Author, I.: Article title. Journal Title-Abbreviated {\bf Vol}, pp--pp (year)
%\bibitem{Ref2}
%Author, I., Smith, J.: Book Title. Publisher, Place (year)
% Format for proceedings
%\bibitem{Ref3}
%Author, I., Smith, J.: Paper title. In: Editor, A. (ed.) Proceedings
%Title, Location, Date, pages. Publisher, Place (year)
% etc
%\end{thebibliography}

\end{document}